\documentstyle[seceq,preprint]{ptptex}

\newcommand{\tr}{\mbox{\rm tr}}
\newcommand{\La}{\mbox{\boldmath$\Lambda$}}
\newcommand{\bl}{\langle\mbox{\boldmath$\Lambda$}|}
\newcommand{\kl}{|\mbox{\boldmath$\Lambda$}\rangle}
\newcommand{\m}{\mbox{\boldmath$\mu$}}
\newcommand{\bm}{\langle\mbox{\boldmath$\mu$}(n)|}
\newcommand{\km}{|\mbox{\boldmath$\mu$}(n)\rangle}
\newcommand{\R}{\mbox{\boldmath$R$}}
\newcommand{\h}{\mbox{\boldmath$H$}}
\newcommand{\th}{\mbox{\boldmath$\theta$}}
\newcommand{\A}{A_\mu^{g^{-1}}}
\newcommand{\J}{{\mathcal J}}

\title{Non-Abelian Stokes Theorem for Wilson Loops Associated with General Gauge Groups}

\author{Minoru {\sc Hirayama} and Masataka {\sc Ueno}}

\inst{Department of Physics, Toyama University, Toyama 930-8555, Japan}

\recdate{July 3, 1999}

\notypesetlogo

\abst{A formula constituting the non-Abelian Stokes theorem for general
semi-simple compact gauge groups is presented.
The formula involves a path integral over a group space
and is applicable to Wilson loop variables irrespective
of the topology of loops.
Some simple expressions analogous to the 't Hooft tensor of a magnetic
monopole are given for the 2-form of interest.
A special property in the case of the fundamental
representation of the group $SU(N)$ is pointed out.}

\begin{document}

\maketitle

\section{Introduction}

The non-Abelian Stokes theorem (NAST) equates the Wilson loop variable
\begin{equation}
W[\gamma]=\tr\bigl(Pe^{i\lambda\int_0^1A_\mu(x(t))
\frac{dx^\mu(t)}{dt}dt}\bigr)
\end{equation}
with a quantity described by the surface integral of the field strength,
\begin{equation}
F_{\mu\nu}(x)=\partial_\mu A_\nu(x)-\partial_\nu A_\mu(x)
-i\lambda[A_\mu(x),A_\nu(x)].
\end{equation}
In (1$\cdot$1), $\gamma$ is a closed loop parametrized by $t$
$(\gamma=\{x(t)$ $|$ $0\leq t\leq1,$ $x(0)=x(1)\})$, $\lambda$ is a gauge
coupling constant, $P$ denotes a path ordering,
and $A_\mu(x)$ is a non-Abelian gauge field.
Roughly speaking, there have been proposed two kinds of NAST.
The first kind of NAST\cite{rf:1}\tocite{rf:13} has a rather long history
and involves a duplicate path ordering.
When the loop $\gamma$ is trivial, i.e. unknotted and unlinked,
it is described as
\begin{eqnarray}
W[\gamma] & = & \tr\bigl(P_te^{i\lambda\int_0^1dt\int_0^1ds
{\mathcal F}_{\mu\nu}(x(s,t))\frac{\partial x^\mu(s,t)}{\partial s}
\frac{\partial x^\nu(s,t)}{\partial t}}\bigr),\\
{\mathcal F}_{\mu\nu}(x) & = & w(x)F_{\mu\nu}(x)w^{-1}(x),
\end{eqnarray}
where $P_t$ denotes the $t$-ordering, $w(x)$ is a unitary matrix depending
on the path from $x(0,0)$ to $x(s,t)$, and the boundary $\partial S$ of the
simply connected surface $S=\{x(s,t)$ $|$ $0\leq s,t\leq1\}$ is assumed to coincide
with $\gamma$.
The point $x(0,0)$ in (1$\cdot$3) should be identical with the
point $x(0)=x(1)$ in (1$\cdot$1). 
Since the matrix $w(x)$ contains the ordering
for a path connecting $x(0,0)$ and $x(s,t)$,
the r.h.s. of (1$\cdot$3) involves a duplicate path ordering.
The consistency of the formula (1$\cdot$3) is guaranteed\cite{rf:10}
by the Bianchi identity,
\begin{equation}
[D_\rho,F_{\mu\nu}]+[D_\mu,F_{\nu\rho}]+[D_\nu,F_{\rho\mu}]=0.
\end{equation}
Generalization of (1$\cdot$3) to the cases that the loop $\gamma$ is
topologically nontrivial, i.e. knotted and/or linked,
was realized recently.\cite{rf:11}

For the second kind of NAST, the path ordering is replaced
with a path integral over a group manifold.\cite{rf:14}\tocite{rf:20}
In the formulation of Diakonov and Petrov, it takes the form\cite{rf:14}
\begin{equation}
W[\gamma]=\int[dg]_\gamma e^{i\int_\gamma \zeta(x)},
\end{equation}
where $\zeta(x)$ is the 1-form defined by
\begin{eqnarray}
\zeta(x) & = & \lambda\bl A_\mu^{g^{-1}}(x)\kl dx^\mu,\\
\A(x) & = & g^{-1}(x)A_\mu(x)g(x)
+\frac{i}{\lambda}g^{-1}(x)\partial_\mu g(x).
\end{eqnarray}
In (1$\cdot$7) and (1$\cdot$8), $g(x)$ is a unitary representation of the
gauge group, $\kl$ is the highest weight state in the representation
and satisfies the normalization condition $\bl\La\rangle=1$.
It can be shown that there exists a group-invariant measure $dg$
satisfying the condition\cite{rf:21}
\begin{equation}
\int dg\hspace{2mm}g\kl\bl g^{-1}=1.
\end{equation}
The integral measure $[dg]_\gamma$ in (1$\cdot$6) is given by
\begin{equation}
[dg]_\gamma=\prod_{x\in\gamma}dg(x).
\end{equation}
It is known that the integral over the group $G$ can be replaced by an
integral over the quotient group $G/H$, where $H$ is the stability group
of $\kl$ defined by
\begin{equation}
h\kl=e^{i\varphi(h)}\kl,\hspace{3mm}
\varphi(h)\in\mbox{\boldmath$R$},\hspace{3mm}h\in H.
\end{equation}
With the aid of the conventional Stokes theorem,
\begin{equation}
\int_\gamma\zeta(x)=\int_Sd\zeta(x),\hspace{5mm}\partial S=\gamma,
\end{equation}
the integrand of the exponent on the r.h.s. of (1$\cdot$6) can be expressed
as a surface integral of a function containing the field strength.
For the fundamental representation of the gauge group $SU(2)$,
Diakonov and Petrov\cite{rf:17} observed
that the 2-form $d\zeta(x)$ in (1$\cdot$12) can be described by a gauge
invariant tensor field which has a structure similar to the
't Hooft tensor.\cite{rf:22}
The latter tensor field is given by
\begin{equation}
T_{\mu\nu}(x)=\sum_{a=1}^3F_{\mu\nu}^a(x)\widehat{\phi}^a(x)
-\frac{1}{\lambda}\sum_{a,b,c=1}^3\epsilon^{abc}\widehat{\phi}^a(x)
(D_\mu\widehat{\phi}(x))^b(D_\nu\widehat{\phi}(x))^c,
\end{equation}
where $\widehat{\phi}^a(x)$ $(a=1,2,3)$ is a unit isovector field
constructed with the $SO(3)$ Higgs field $\phi^a(x)$ as
$\widehat{\phi}^a(x)=\phi^a(x)/|\mbox{\boldmath$\phi$}(x)|$.
It is well known that the electromagnetic field produced by the $SO(3)$
magnetic monopole is given by $T_{\mu\nu}(x)$.
If we replace $\widehat{\phi}^a(x)$ in (1$\cdot$13) by
\begin{equation}
n^a(x)=\frac{1}{2}\tr(g^{-1}(x)\tau^ag(x)\tau^3),
\end{equation}
with $\tau^a$ $(a=1,2,3)$ being Pauli matrices, we obtain the
tensor field appearing in $d\zeta(x)$.

The purpose of this paper is to pursue the line of thought of the second kind of
NAST\cite{rf:14,rf:17} further.
We first discuss the case of the general semi-simple compact gauge groups.
We find some interesting expressions for the integrand $d\zeta(x)$
of the above-mentioned surface integral.
We also find that there exists a simpler expression for $d\zeta(x)$
if we restrict ourselves to a fundamental representation of $SU(N)$.
All the above expressions for $d\zeta(x)$ involve natural generalizations
of the 't Hooft tensor.

This paper is organized as follows. In \S2 we obtain some expressions for the
2-form $d\zeta(x)$ which can be used for general semi-simple compact gauge groups.
In \S3 we discuss the case of the fundamental representation of
$SU(N)$ and clarify its special property.
Section 4 is devoted to summary and discussion.

\section{Some expressions for the integrand of the surface integral}

\subsection{Preliminaries}

We consider a fixed $D$-dimensional representation of a semi-simple compact
Lie group $G$. The representations
$\{T^a$ $|$ $a=1,2,\cdots,k\}=\{H_i,E_\alpha$ $|$ $i=1,2,\cdots,l,$
$\alpha$: root of $G\}$ of the generators of $G$ are assumed to satisfy
\begin{eqnarray}
[H_i,H_j] & = & 0,\hspace{12mm}i,j=1,2,\cdots,l,\nonumber\\
{}[H_i,E_\alpha] & = & \alpha_iE_\alpha,\hspace{8.8mm}i=1,2,\cdots,l,\nonumber\\
\tr(T^aT^b) & = & \kappa\delta^{ab},\hspace{6.1mm}a,b=1,2,\cdots,k,
\end{eqnarray}
where $\alpha=(\alpha_1,\alpha_2,\cdots,\alpha_l)\in\R^l$
is a root vector and $\kappa$ is a positive constant.
Thus, we are working in a $D$-dimensional representation of a group of
dimension $k$ and rank $l$.
The basis vectors $\km$ $(n=1,2,\cdots,D)$ of the representation space satisfy
\begin{eqnarray}
H_i\km & = & \mu_i(n)\km,\hspace{3mm}i=1,2,\cdots,l,\\
\bm\m(m)\rangle & = & \delta_{mn},\hspace{3mm}
\sum_{n=1}^D\km\bm=1,
\end{eqnarray}
where $\m(n)=(\mu_1(n),\mu_2(n),\cdots,\mu_l(n))\in\R^l$ is a weight vector.
We denote the highest weight of the representation and
the highest weight state by $\La$ and $\kl$, respectively:
\begin{eqnarray}
\La & = & \m(1)=(\Lambda_1,\Lambda_2,\cdots,\Lambda_l),\\
\kl & = & |\m(1)\rangle.
\end{eqnarray}
We define a group-theoretic coherent state $|g\rangle$ by
\begin{equation}
|g\rangle=g\kl,
\end{equation}
where $g$ is a $D$-dimensional representation of an element of $G$.
Then, as discussed by many authors, the Wilson loop variable defined by
(1$\cdot$1) can be written as (1$\cdot$6).\cite{rf:14}\tocite{rf:21,rf:23}
If we make use of (1$\cdot$6) and (1$\cdot$12), we are led to the formula
\begin{equation}
W[\gamma]=\int[dg]_\gamma e^{i\int_S\omega},
\end{equation}
where $\omega$ is given by
\begin{equation}
\omega=d\zeta(x)=\lambda d\bigl(\bl\A(x)\kl dx^\mu\bigr).
\end{equation}

\subsection{A formula for $\bl K\kl$}

Diakonov and Petrov\cite{rf:14} argued that the 1-form $\zeta(x)$ in (1$\cdot$7)
is given by the trace of the product of the quantity
$\La\cdot\h=\sum_{i=1}^l\Lambda_iH_i$
and $\A(x)$ in (1$\cdot$8).
We begin our discussion by explicitly deriving their conclusion.
If we define diagonal matrices $J_n$ $(n=1,2,\cdots,D)$ by
\begin{equation}
\km\bm=\frac{1}{\kappa}\m(n)\cdot\h+J_n,
\end{equation}
it can be shown that, they satisfy
\begin{equation}
\tr(H_iJ_n)=\tr(E_\alpha J_n)=0.
\end{equation}
Then, we have
\begin{eqnarray}
\bl K\kl & = & \tr(\kl\bl K)\nonumber\\
& = & \tr\Bigl(\Bigl\{\frac{1}{\kappa}\La\cdot\h+J_1\Bigr\}K\Bigr)\nonumber\\
& = & \frac{1}{\kappa}\tr(\La\cdot\h K),\hspace{3mm}K\in{\mathcal G}_D,
\end{eqnarray}
where ${\mathcal G}_D$ is a $D$-dimensional representation of the Lie algebra of $G$,
i.e. a linear span of the $H_i$ and the $E_\alpha$.
The formula (2$\cdot$11) reproduces the conclusion of Diakonov and Petrov.
We note that the formula
\begin{equation}
H_i=\sum_{n=1}^D\mu_i(n)\km\bm
\end{equation}
yields the relations
\begin{eqnarray}
& \sum_{n=1}^D & \mu_i(n)\mu_j(n)=\kappa\delta_{ij},\nonumber\\
& \sum_{n=1}^D & \mu_i(n)J_n=0
\end{eqnarray}
and the expression
\begin{equation}
J_n=\sum_{m=1}^D\Bigl\{\delta_{nm}
-\frac{1}{\kappa}\m(n)\cdot\m(m)\Bigr\}|\m(m)\rangle\langle\m(m)|.
\end{equation}

\subsection{An Abelian-like expression for $\omega$} 

From (2$\cdot$8) and (2$\cdot$11), we obtain the formula
\begin{equation}
\omega=\frac{\lambda}{\kappa}\tr\bigl((\La\cdot\h)d(A_\mu^{g^{-1}}(x)dx^\mu)\bigr).
\end{equation}
If we define a vector field $B_\mu(x)$ by
\begin{equation}
B_\mu(x)=\frac{1}{\kappa}\tr\bigl((\La\cdot\h)(A_\mu^{g^{-1}}(x))\bigr),
\end{equation}
$\omega$ becomes
\begin{eqnarray}
\omega & = & \lambda G_{\mu\nu}(x)d\sigma^{\mu\nu},\\
G_{\mu\nu}(x) & = & \partial_\mu B_\nu(x)-\partial_\nu B_\mu(x),\\
d\sigma^{\mu\nu} & = & \frac{1}{2}dx^\mu\wedge dx^\nu.
\end{eqnarray}
We can interpret the factor $\int_S\omega$ in (2$\cdot$7) as the flux of the field
strength produced by the potential $B_\mu(x)$.
It should be noted that the transformation property of $B_\mu(x)$ under
$\A(x)\rightarrow(A_\mu^{g'})^{g^{-1}}(x)=A_\mu^{g^{-1}g'}(x)$
is not simple and that, under the transformation
\begin{equation}
(\A)(x)\rightarrow(A_\mu^{g^{-1}})^{g'}(x)=A_\mu^{g'g^{-1}}(x)
\end{equation}
with
\begin{equation}
g'(x)=\exp[i\th(x)\cdot\h],\hspace{5mm}\theta_i(x)\in\R,
\hspace{5mm}(i=1,2,\cdots,l)
\end{equation}
$B_\mu(x)$ transforms in the following way:
\begin{equation}
B_\mu(x)\rightarrow B_\mu(x)
+\frac{1}{\lambda}\partial_\mu(\La\cdot\th(x)).
\end{equation}

\subsection{A 't Hooft tensor-like expression for $\omega$}

To obtain another expression for $\omega$, we separate $\A(x)$ into
two terms as in (1$\cdot$7). Then we have
\begin{eqnarray}
\omega & = & \xi-\eta,\\
\xi & = & \frac{1}{\kappa}\sum_{i=1}^l\Lambda_i\xi_i,
\hspace{3mm}\eta=\frac{1}{\kappa}\sum_{i=1}^l\Lambda_i\eta_i,\\
\xi_i & = & \lambda\tr\bigl(H_id(g^{-1}(x)A_\mu(x)g(x))dx^\mu\bigr),\\
\eta_i & = & i\tr\bigl(H_ig^{-1}(dg)g^{-1}dg\bigr).
\end{eqnarray}
The 2-form $\eta_i$ is called the Kirillov-Kostant 2-form.
It is known\cite{rf:24} that, making use of $Q(g)$ defined by
\begin{equation}
Q(g)=g(x)\kl\bl g^{-1}(x)
=\frac{1}{\kappa}g(x)\La\cdot\h g^{-1}(x)
+g(x)J_1g^{-1}(x),
\end{equation}
the 2-form $\eta$ can be written as
\begin{equation}
\eta=-i\tr\bigl(Q(g)[\partial_\mu Q(g),\partial_\nu Q(g)]\bigr)d\sigma^{\mu\nu}.
\end{equation}
The first term $\xi$ on the r. h. s. of (2$\cdot$23) can be rewritten as
$$\xi=\lambda\tr\bigl(Q(g)[F_{\mu\nu}(x)+i\lambda[A_\mu(x),A_\nu(x)]
-2\{A_\mu(x),\partial_\nu Q(g)\}]\bigr)d\sigma^{\mu\nu}.$$
After some manipulations, it turns out that $G_{\mu\nu}(x)$ in (2$\cdot$17)
can be expressed as
\begin{eqnarray}
G_{\mu\nu}(x) & = & \tr\Bigl(Q(g)\Bigl\{F_{\mu\nu}(x)
+\frac{i}{\lambda}[D_\mu Q(g),D_\nu Q(g)]\Bigr\}\Bigr),\nonumber\\
D_\mu Q(g) & = & \partial_\mu Q(g)-i\lambda[A_\mu(x),Q(g)].
\end{eqnarray}
The similarity of $G_{\mu\nu}(x)$ to $T_{\mu\nu}(x)$ of (1$\cdot$11) is obvious.
We note that the formulas
\begin{equation}
\{Q(g)\}^2=Q(g)
\end{equation}
and
\begin{equation}
Q(g)\{\partial_\mu Q(g)\}Q(g)=0
\end{equation}
were repeatedly utilized in the derivation of (2$\cdot$29).
We also note that the operator $Q(g)$ appearing in $G_{\mu\nu}(x)$
does not belong to ${\mathcal G}_D$ of (2$\cdot$11) since
it contains $g(x)J_1g^{-1}(x)$, while $M_i(g)$ defined by
\begin{equation}
M_i(g)=g(x)H_ig^{-1}(x),\hspace{3mm}i=1,2,\cdots,l,
\end{equation}
belongs to ${\mathcal G}_D$.

\subsection{The third expression for $\omega$}

In terms of $M_i(g)$ introduced in the last subsection, the tensor field
$G_{\mu\nu}(x)$ can be written as
\begin{equation}
G_{\mu\nu}(x)=\frac{1}{\kappa}\tr\bigl((\La\cdot\mbox{\boldmath$M$})
\{F_{\mu\nu}(x)+H_{\mu\nu}(x)\}\bigr),
\end{equation}
where $H_{\mu\nu}(x)\in{\mathcal G}_D$ is defined by
\begin{eqnarray}
H_{\mu\nu}(x) & = & \frac{i}{\lambda}[K_\mu(x),K_\nu(x)],\\
K_\mu(x) & = & \lambda A_\mu(x)+i(\partial_\mu g(x))g^{-1}(x).
\end{eqnarray}
It can be readily seen that $K_\mu(x)$ and $H_{\mu\nu}(x)$
transform covariantly under gauge transformations.
In the next section, we consider the special case of the fundamental
representation of $SU(N)$. 
There, we find an expression for $G_{\mu\nu}(x)$ 
which is also similar to (2$\cdot$29) and (1$\cdot$13).

\section{The case of the fundamental representation of $SU(N)$}

In the previous section, we discussed the path integral formulation of the NAST
by making use of group-theoretic coherent states.
It turned out that the integrand $\omega$ of the surface integral,
or $G_{\mu\nu}(x)$ in (2$\cdot$17), can be expressed by the operator $Q(g)$
of (2$\cdot$27) in a form analogous to the 't Hooft tensor $T_{\mu\nu}(x)$,
(1$\cdot$13), of the $SO(3)$ magnetic monopole.
We stress that $Q(g)$ is not Lie algebra valued because it has the 
piece $g(x)J_1g^{-1}(x)$.
In this section, we show that, in the case of the fundamental representation
of $SU(N)$, we can rewrite $G_{\mu\nu}(x)$ in terms of the Lie algebra
valued quantity $M(g)$ defined by
\begin{equation}
M(g)=\frac{1}{\kappa}g(x)\La\cdot\h g^{-1}(x)
=\frac{1}{\kappa}\sum_{i=1}^{N-1}\Lambda_iM_i(g).
\end{equation}
By definition, $M(g)$ can be written as
\begin{eqnarray}
M(g) & = & \frac{1}{\kappa}\sum_{a=1}^{N^2-1}M^a(g)T^a,\\
M^a(g) & = & \tr(T^aM(g))\nonumber\\
& = & \bl g^{-1}(x)T^ag(x)\kl,
\end{eqnarray}
where $N^2-1$ is the dimension of the group $SU(N)$.
The rank of $SU(N)$ is $N-1$ and hence there are $N-1$
fundamental representations.
The following discussion is applicable to all of them.
Since the dimension of the fundamental representation of $SU(N)$ is $N$,
we are led to the conclusion that all of the
diagonal matrices $J_n$ in (2$\cdot$9) are given by
\begin{equation}
J_n=\frac{1}{N}\mbox{\boldmath$1$},\hspace{3mm}n=1,2,\cdots,N,
\end{equation}
where {\boldmath$1$} is the $N\times N$ unit matrix.
The above conclusion is deduced from the requirements $\tr(H_iJ_n)=0$,
$\tr(H_i)=0$ $(i=1,2,\cdots,N-1)$ and $\tr(\km\bm)=1$.
From (2$\cdot$27), (3$\cdot$1) and (3$\cdot$4), we obtain
\begin{equation}
Q(g)=M(g)+\frac{1}{N}\mbox{\boldmath$1$},
\end{equation}
and hence
\begin{equation}
\partial_\mu Q(g)=\partial_\mu M(g).
\end{equation}
With the help of (2$\cdot$10), (2$\cdot$29), (3$\cdot$5) and (3$\cdot$6),
we easily find the formula
\begin{eqnarray}
G_{\mu\nu}(x) & = & \tr\Bigl(M(g)\Bigl\{F_{\mu\nu}(x)
+\frac{i}{\lambda}[D_\mu M(g),D_\nu M(g)]\Bigr\}\Bigr),\\
D_\mu M(g) & = & \partial_\mu M(g)-i\lambda[A_\mu(x),M(g)].
\end{eqnarray}
In terms of the components $M^a(g)$ in (3$\cdot$2),
the tensor field $G_{\mu\nu}(x)$ is expressed as
\begin{equation}
G_{\mu\nu}(x)=\sum_{a=1}^{N^2-1}F_{\mu\nu}^a(x)M^a(x)
-\frac{1}{\lambda}\sum_{a,b,c=1}^{N^2-1}f^{abc}M^a(g)(D_\mu M(g))^b(D_\nu M(g))^c,
\end{equation}
where $f^{abc}$ is the structure constant of $SU(N)$.
Comparing (3$\cdot$9) with (1$\cdot$13), we observe that there exists
a complete parallelism between the 't Hooft tensor $T_{\mu\nu}(x)$
and the $G_{\mu\nu}(x)$ in the fundamental representation of $SU(N)$.

To end this section, we compare our result with the expression for
$\eta_i$ in (2$\cdot$24) given by Faddeev and Niemi:\cite{rf:26}
\begin{equation}
\eta_i=-i\tr\Bigl(M_i(g)\sum_{k=1}^{N-1}
[\partial_\mu M_k(g),\partial_\nu M_k(g)]\Bigr)d\sigma^{\mu\nu}.
\end{equation}
On the other hand, through analysis similar to the above, we obtain
\begin{equation}
\eta=-i\tr(M(g)[\partial_\mu M(g),\partial_\nu M(g)])d\sigma^{\mu\nu}.
\end{equation}
The fact that the expressions (3$\cdot$10) and (3$\cdot$11) are consistent with
the relation $\eta=(\sum_{i=1}^{N-1}\Lambda_i\eta_i)/\kappa$, (2$\cdot$24),
can be checked in the following way. With the help of the formulas
\begin{eqnarray}
M_i(g)M_j(g) & = & \frac{1}{2}\sum_{k=1}^{N-1}d_{ijk}M_k(g)
+\frac{1}{2N}\delta_{ij}\\
\sum_{i=1}^{N-1}d_{iik} & = & 0,
\end{eqnarray}
with $d_{ijk}\equiv4\tr(H_iH_jH_k)$, which are used in Ref. 26),
we can rewrite (3$\cdot$10) as
\begin{eqnarray}
\eta_i & = & i\tr(M_i(g)[\J_\mu,\J_\nu])d\sigma^{\mu\nu},\\
\J_\mu & = & (\partial_\mu g(x))g^{-1}(x).
\end{eqnarray}
As for $\eta$ in (3$\cdot$11), we make use of the formula
\begin{equation}
\{M(g)\}^2=\frac{N-2}{N}M(g)+\frac{N-1}{N^2}\mbox{\boldmath$1$},
\end{equation}
derived from (2$\cdot$30) and (3$\cdot$5). Then we have
\begin{equation}
\eta=i\tr(M(g)[\J_\mu,\J_\nu])d\sigma^{\mu\nu}.
\end{equation}
Recalling the definitions (2$\cdot$24) and (3$\cdot$1),
it is clear that (3$\cdot$14) and (3$\cdot$17),
and hence (3$\cdot$10) and (3$\cdot$11), are consistent.
It is, of course, possible to rewrite $G_{\mu\nu}(x)$ of (3$\cdot$7)
and (3$\cdot$9) in terms of $M_i(g)$ instead of $M(g)$.
In our opinion, however, the expressions (3$\cdot$7) and (3$\cdot$9) are
simplest.

\section{Summary and discussion}

We have presented some expressions for the 2-form $\omega$ appearing
in the NAST (2$\cdot$7).
It turns out that the simple formulas (2$\cdot$9) and (2$\cdot$10),
which lead us to (2$\cdot$11), are useful.
We hope that the formulas (2$\cdot$18), (2$\cdot$29) and (2$\cdot$33)
for the general case and  (3$\cdot$7) and (3$\cdot$9) for the case of the 
fundamental representation of $SU(N)$ are helpful for future investigations
of the Wilson loop.
The NAST (2$\cdot$7) can be applied to any closed loop $\gamma$, since only
the conventional Stokes theorem, (1$\cdot$11), has been used in the derivation.
In contrast to Abelian cases,
the expressions (2$\cdot$29) and (3$\cdot$9) for $G_{\mu\nu}(x)$
still contain the gauge potential
$A_\mu(x)$ through the covariant derivative $D_\mu$.
The formula (2$\cdot$7) is, in a sense, of a curious structure:
the integral measure $[dg]_\gamma$ concerns the loop $\gamma=\partial S$,
while the exponent $\int\omega$ concerns
the interior $S^0$ of $S$ as well as $\gamma$.
If necessary, we can insert the path integral 
$\int[\widetilde{d}g]_{S^0}=1$ over the interior $S^0$ with
$[\widetilde{d}g]_{S^0}=\prod_{x\in S^0}\widetilde{d}g(x)$,
$\widetilde{d}g(x)=|\bl g(x)\kl|^2dg(x)$ into (1$\cdot$6), since
we obtain $\int\widetilde{d}g(x)=1$, $^\forall x$, from (1$\cdot$8).
If we make use of (1$\cdot$12) after the insertion,
we are led to the formula
$W[\gamma]=\int[dg]_\gamma[\widetilde{d}g]_{S^0}e^{i\int_S\omega}$.

\section*{Acknowledgements}

The authors are grateful to Hitoshi Yamakoshi for discussions.

\end{document}